\documentclass{article}
\usepackage{ciss2005}
\usepackage{graphicx, amssymb}
\usepackage{graphicx}
\newtheorem{thm}{Theorem}[section]
\newtheorem{lem}[thm]{Lemma}

\newtheorem{prop}[thm]{Proposition}

\begin{document}

\itwtitle       {Analysis of Second-order Statistics Based
Semi-blind Channel Estimation in CDMA Channels\\}

\itwauthor      {Husheng Li and H. Vincent Poor\footnotemark[1]}
                {Department of Electrical Engineering \\
         Princeton University \\
         Princeton, NJ 08544 \\
         e-mail: {\tt \{hushengl,poor\}@princeton.edu}}

%

\itwmaketitle

\footnotetext[1]{This research was supported in part by the Office
of Naval Research under Grant N00014-03-1-0102 and in part by the
New Jersey Center for Wireless Telecommunications.}

\begin{itwabstract}
The performance of second order statistics (SOS) based semi-blind
channel estimation in long-code DS-CDMA systems is analyzed. The
covariance matrix of SOS estimates is obtained in the large system
limit, and is used to analyze the large-sample performance of two
SOS based semi-blind channel estimation algorithms. A notion of
blind estimation efficiency is also defined and is examined via
simulation results.
\end{itwabstract}

\begin{itwpaper}

\itwsection{Introduction}

In practical wireless communication systems, the transmitted signals
usually experience fading, which either attenuates the received
power or causes dispersion. Usually, the channel state information
is unknown to both the transmitter and the receiver in many
practical applications, thus necessitating channel estimation at the
frontend of the coherent receiver. In this paper, we focus on the
long-code direct-sequence code division multiple-access (DS-CDMA)
systems and frequency-selective fading channels.

A variety of channel estimation algorithms have been derived, based
primarily on one of two aspects of random observations, namely, the
distribution and the moments. When the distribution of the received
signal conditioned on the channel state information is known, the
maximum likelihood (ML) criterion can be applied to yield
asymptotically optimal performance. Such ML channel estimation
algorithms \cite{Buzzi2001} are suitable for training-symbol-based
systems, in which a fraction of the transmitted symbols is known to
both the transmitter and receiver. However, the channel state
information hidden in the information symbols is ignored, thus
decreasing the spectral efficiency.

When only the information symbols are available (usually called
\textit{blind} channel estimation), the moment based channel
estimation algorithms require substantially less computational cost
than other methods, thus being feasible for practical applications.
Thus, over the past two decades, a large number of blind channel
estimation algorithms
\cite{Bensley1996}\cite{Liu1996}\cite{ZhengyuanXu2000}\cite{Zeng1997}
have been developed using moment estimation, particularly the second
order statistics (SOS). Typically, based on SOS estimation, the
subspace method \cite{Bensley1996} and moment matching
~\cite{ZhengyuanXu2000} can be applied. The subspace technique is
suitable only for stationary channels, e.g. short-code CDMA systems,
since it requires the signal subspace to be time-invariant. However,
in many practical CDMA systems, long codes are employed, thus making
the overall channel non-stationary. Therefore, only the moment
matching technique can be applied directly to long-code CDMA
systems. When training symbols are available, both the training and
information symbols can be used jointly to yield better estimates
(usually called \textit{semi-blind} channel estimation).

In this paper, we adopt the moment-matching-based algorithm of
\cite{ZhengyuanXu2000} for estimating the SOS of channel
coefficients of each user. The following two SOS based semi-blind
channel estimation algorithms are considered.
\begin{itemize}
\item \textbf{Moment-matching-based estimation}. The
asymptotically optimal moment based estimator \cite{Porat1993} is
applied to the first order moment, estimated from the training
symbols, and the SOS, estimated from the information symbols, thus
resulting in moment-matching-based semi-blind channel estimation
under some assumptions.

\item \textbf{Subspace-based estimation}. The subspace spanned by
the channel coefficient vector is obtained from the eigenstructure
of SOS estimation. The channel estimates are obtained by projecting
the training symbols onto the subspace followed by linear combining.
It should be noted that the \textit{subspace} here is different from
that mentioned above, which is spanned by the signal.
\end{itemize}

The remainder of this paper is organized as follows. A discrete
model of the frequency selective fading channels is explained in
Section II. The performance of SOS estimation and semi-blind channel
estimation is analyzed in Section III and Section IV, respectively.
Simulation results and conclusions are given in Section V and
Section VI, respectively. The details can be found in the journal
version of this paper \cite{LiHuPreprint}.

\itwsection{System Model} We consider a synchronous uplink DS-CDMA
system operating over frequency selective fading channels of order
$P$ (i.e, $P$ is the delay spread in chip intervals). Let $K$ denote
the number of active users, $N$ the spreading gain and
$\beta\triangleq\frac{K}{N}$ the system load. We model the frequency
selective fading channels with discrete finite-impulse-response
(FIR) filters. The $z$-transform of the channel response of user $k$
is thus given by
\begin{eqnarray}
h_k(z)=\sum_{p=0}^{P-1}g_k(p)z^p,
\end{eqnarray}
where $\{g_k(p)\}_{p=0,...,P-1}$ are the corresponding independent
and identically distributed (i.i.d.) (with respect to both $k$ and
$p$) channel coefficients with variance $\frac{1}{P}$. We adopt the
block fading model, in which the channel coefficients, namely
$\{{g}_k(p)\}$, remain unchanged during a coherent block, and we
denote the coherence time by $M$, measured in channel symbol
periods.

With the assumption $P\ll N$, we can ignore the effects of
intersymbol interference (ISI). Thus, the chip matched filter output
at the $l$-th chip period in the $m$-th symbol period can be written
as
\begin{eqnarray}\label{rec_signal}
r(mN+l)=\sum_{k=1}^Kx_k(m)h_k^{(m)}(l)+n(mN+l),
\end{eqnarray}
where $\{x_k(m)\}$ denotes the complex channel symbols, which are
mutually independent with respect to $k$ and $m$. We assume that
$x_k(m)$ is a complex random variable satisfying
$E\left\{\left|x_k(m)\right|^2\right\}=1$ and
$E\left\{x_k^2(m)\right\}=0$, which hold, for example, for quadratic
phase shift keying (QPSK) modulation. $\{n(mN+l)\}$ is additive
white complex Gaussian noise, which satisfies
$E\{|n(mN+l)|^2\}=\sigma_n^2$ \footnote{Note that $\sigma_n^2$ is
the noise variance, normalized to represent the inverse
signal-to-noise ratio.}. $\{h_k^{(m)}(l)\}$ denote the convolution
of the spreading code and the channel coefficients:
\begin{eqnarray}
h_k^{(m)}(l)=s_k^{(m)}(l)\star g_k(l),
\end{eqnarray}
where $s_k^{(m)}(l)$ is the $l$-th chip of the original random
spreading codes of user $k$ in symbol period $m$, which takes values
$\frac{1}{\sqrt{N}}$ and $-\frac{1}{\sqrt{N}}$ equiprobably and
independently  with respect to $k$, $m$ and $l$.

When the channel coefficients are unknown to the receiver, channel
estimation is necessary for coherent reception. For simplicity, we
assume that $M_t$ training symbols, which are known to both
transmitter and receiver, are transmitted at the beginning of each
coherent block, and the proportion of training symbols is denoted by
$\alpha=\frac{M_t}{M}$.

\itwsection{Performance of SOS Estimation} When discussing SOS
estimation in this section, for notational simplicity, we assume
that all $M$ channel symbols during the coherent block are unknown
to the receiver. When analyzing the performance of semi-blind
channel estimation, we can simply replace $M$ with $M-M_t$ in the
conclusions for SOS estimation. It should be noted that our analysis
on the performance of SOS estimation is based on the large system
limit, namely $K,N\rightarrow\infty$ while keeping $\beta$ and $P$
constant.

\subsection{SOS Estimation Algorithm} We adopt the algorithm
proposed in ~\cite{ZhengyuanXu2000} to estimate the channel
coefficients in long-code CDMA systems. On
defining\footnote{Superscript $T$ denotes transposition and
superscript $H$ denotes conjugate transposition.}
$\mathbf{r}(m)=\left(r(mN+P),...,r((m+1)N)\right)^T$ as the signal
unaffected by ISI during the symbol period $m$, we can rewrite
(\ref{rec_signal}) in vector form, which is given by
\begin{eqnarray}\label{SignalModel}
   \mathbf{r}(m)=\sum_{k=1}^KC_k^{(m)}\mathbf{g}_kx_k(m)+\mathbf{n}(m),
\end{eqnarray}
where $\mathbf{g}_k=\left(g_k(0),...,g_k(P-1)\right)^T$,
$\mathbf{n}_k(m)=\left(n(mN+P),...,n((m+1)N)\right)^T$ and
$C_k^{(m)}$ is the truncated Sylvester matrix, which is given by
\begin{scriptsize}
\begin{eqnarray}
C_k^{(m)}=\left(\begin{array}{llll}s_k^{(m)}(P) & s_k^{(m)}(P-1) & \ldots & s_k^{(m)}(1)\\
                         s_k^{(m)}(P+1) & s_k^{(m)}(P) & \ldots & s_k^{(m)}(2)\\
                         \vdots & \vdots &\ldots  &\vdots  \\
                         s_k^{(m)}(N) & s_k^{(m)}(N-1) & \ldots & s_k^{(m)}(N-P+1)\\
                         \end{array}\right)\nonumber.
\end{eqnarray}
\end{scriptsize}

We define a cost function for moment matching, which is given by
\begin{eqnarray}
  J=\left\|\sum_{m=1}^M\left(E\left\{\mathbf{R}_r(m)\right\}-\mathbf{R}_r(m)\right)\right\|_F,
\end{eqnarray}
where $\mathbf{R}_r(m)=\mathbf{r}(m)\mathbf{r}(m)^H$ and
$\|\cdot\|_F$ denotes the Frobenius norm. The optimal SOS is
obtained by minimizing the cost function $J$,
$$
\hat{\mathbf{d}}=\arg\min_d J,
$$
where $\mathbf{d}=\left(\mathbf{d}_1^H,...,\mathbf{d}_k^H\right)^H$,
$\mathbf{d}_k=\mbox{vec}\left(\mathbf{g}_k\mathbf{g}_k^H\right)$ and
$\mbox{vec}(X)$ denotes a vector formed by stacking the columns of
matrix $X$ into one column vector. An explicit expression for
$\hat{\mathbf{d}}$ is obtained in ~\cite{ZhengyuanXu2000}, which is
given by
\begin{eqnarray}\label{7}
   \hat{\mathbf{d}}=\mathbf{T}^{-1}\mathbf{y},
\end{eqnarray}
where \footnote{The dimensions of the matrices are labeled as
subscripts.}
$$
\mathbf{T}_{P^2K\times P^2K}=\frac{1}{M}\sum_{m=1}^{M}Q^T(m)Q(m),
$$
$$
\mathbf{y}_{P^2K\times
1}=\frac{1}{M}\sum_{m=1}^MQ^T(m)\mbox{vec}\left(\mathbf{r}(m)\mathbf{r}^H(m)\right)-\sigma_n^2Q^T(m)\mbox{vec}(\mathbf{I}),
$$
$$
Q(m)_{(N-P+1)^2\times P^2K}=\left(Q_1(m),Q_2(m),...,Q_K(m)\right),
$$
and
$$
Q_k(m)_{(N-P+1)^2\times P^2}=C_k^{(m)}\otimes C_k^{(m)},
$$
where $\otimes$ denotes the Kronecker product. In practical
implementations, $\hat{\mathbf{d}}$ can be obtained by solving
(\ref{7}) iteratively.

\subsection{Identifiability and Consistency} On analyzing the elements of $\mathbf{T}$, we obtain the following lemma.

\begin{lem}\label{LemmaIII.2}
$\mathbf{T}$ converges elementwise to $\mathbf{I}$ almost surely as
$M\rightarrow \infty$ and $N\rightarrow \infty$.
\end{lem}

In ~\cite{ZhengyuanXu2000}, it is shown that the channel is
identifiable and that SOS based channel estimation is consistent if
$\mathbf{T}$ is nonsingular. We cannot draw the conclusion that
$\mathbf{T}$ is nonsingular as $M\rightarrow\infty$ and
$N\rightarrow\infty$ merely from Lemma \ref{LemmaIII.2}, since the
corresponding convergence is elementwise and not in norm. However,
we can show the following proposition, which states a sufficient
condition for the non-singularity of $E\{\mathbf{T}\}$.
\begin{prop}
Suppose $K$ and $N$ are such that $KP < N - P + 1$.  Then
$E\{\mathbf{T}\}$ is nonsingular  and
$\left\|E\{\mathbf{T}\}^{-1}-\mathbf{I}\right\|_1$ is
finite\footnote{$\|A_{n\times n}\|_1\triangleq \max_{1\leq j\leq
n}\sum_{i=1}^n\left|A_{ij}\right|$.}.
\end{prop}

As $K$ and $N$ become sufficiently large, the condition that
$KP<N-P+1$ is equivalent to $\beta<\frac{1}{P}$. Since $\mathbf{T}$
converges to $E\{\mathbf{T}\}$ almost surely as
$M\rightarrow\infty$, $\mathbf{T}$ is nonsingular with large
probability when $M$ is sufficiently large. Since
$\left\|E\{\mathbf{T}\}^{-1}-\mathbf{I}\right\|_1$ is finite as
$K,N\rightarrow\infty$, most of the elements of
$E\{\mathbf{T}\}^{-1}$ converge to the corresponding elements of
$\mathbf{I}$. Thus it is reasonable to approximate $\mathbf{T}^{-1}$
with $\mathbf{I}$ when $K$ and $N$ are sufficiently large. For the
case $\beta>\frac{1}{P}$, numerical simulation indicates that the
non-singularity of $\mathbf{T}$ still typically holds. Thus, in the
following analysis of SOS estimation error, we assume that
$\mathbf{T}=\mathbf{I}$, which will be validated later by simulation
results.

\subsection{SOS Estimation Error} The SOS estimation error $\delta
\mathbf{d}_k\triangleq \hat{\mathbf{d}}_k-\mathbf{d}_k$ is due to
the multiple-access interference (MAI) and noise. The following
proposition gives an explicit expression for the covariance matrix
of $\delta \mathbf{d}_k$, where the assumptions can be shown to hold
in elementwise convergence as $K,N\rightarrow\infty$.

\begin{prop}\label{PropSOSMSE}
Assume
\begin{eqnarray}
\sum_{i,j=1,i\neq
j}^KQ_{ij}(m)\mathbf{d}_{ij}\mathbf{d}_{ij}^HQ_{ij}^T(m)=\beta^2\mathbf{I}_{(N-P+1)^2\times(N-P+1)^2}.\nonumber
\end{eqnarray}
$$
\sum_{k=2}^K\left(C_k^{(m)}\otimes
\mathbf{n}(m)\right)\mathbf{g}^*_k\mathbf{g}^T_k\left(C_k^{(m)}\otimes
\mathbf{n}(m)\right)^H=\beta\sigma_n^2
$$
and
$$
\sum_{k=1}^K\left(\mathbf{n}^*(m)\otimes
C_k^{(m)}\right)\mathbf{g}_k\mathbf{g}^H_k\left(\mathbf{n}^*(m)\otimes
C_k^{(m)} \right)^H=\beta\sigma_n^2.
$$

Then we have, for $k=1,...,K$, as $K,N\rightarrow\infty$,
\begin{small}
\begin{eqnarray}\label{15}
\Sigma^k_{\delta\mathbf{d}}&\triangleq&
ME\left\{\delta\mathbf{d}_k\delta\mathbf{d}_k^H\right\}\nonumber\\
&\rightarrow&
\left(2\beta\sigma_n^2+(\sigma_n^2)^2+\beta^2+\frac{2\beta}{P}\right)\mathbf{I}_{P^2\times
P^2}+\sigma_n^2{\bf{\Omega}}_k.
\end{eqnarray}
\end{small}
where
\begin{scriptsize}
\begin{eqnarray}
\left(\bf{\Omega}_k\right)_{ij}=\left\{
\begin{array}{llll}
\left|\left(\mathbf{g}_k\right)_{\left\lceil\frac{i}{P}\right\rceil}\right|^2+\left|\left(\mathbf{g}_k\right)_{\mbox{mod}(i,P)}\right|^2,\qquad\mbox{if
}i=j\\
\left(\mathbf{g}_k\right)^*_{\left\lceil\frac{i}{P}\right\rceil}\left(\mathbf{g}_k\right)_{\left\lceil\frac{j}{P}\right\rceil},\qquad\mbox{otherwise,
if
}\mbox{mod}(i,P)=\mbox{mod}(j,P)\\
\left(\mathbf{g}_k\right)_{\mbox{mod}(i,P)}\left(\mathbf{g}_k\right)^*_{\mbox{mod}(j,P)},\qquad\mbox{otherwise,
if
}\left\lceil\frac{i}{P}\right\rceil=\left\lceil\frac{j}{P}\right\rceil\\
0,\qquad\mbox{otherwise}
\end{array}
\right.\nonumber.
\end{eqnarray}
\end{scriptsize}
\end{prop}

An interesting observation on (\ref{15}) is that the SOS estimation
errors are mutually correlated and the covariance matrix is
dependent on the realization of channel coefficients. However, when
$\beta$ is moderate or large, the correlation is weak, compared with
the error variance. We can apply the asymptotic results in
(\ref{15}) to finite systems as approximations.

For the average SOS estimation variance of all users, we have
\begin{eqnarray}
\sigma_d^2&\triangleq &
\frac{1}{KP^2}\sum_{k=1}^K\Sigma_{\delta d}^k\nonumber\\
&\rightarrow&2\beta\sigma_n^2+\left(\sigma_n^2\right)^2+\beta^2+\frac{2(\beta+\sigma_n^2)}{P},\nonumber
\end{eqnarray}
from which we can see that the performance of SOS estimation is
determined by the system load and noise variance.

It should be noted that it is not sufficient to obtain
$E\left\{\delta\mathbf{d}_k\delta\mathbf{d}_k^H\right\}$ for the
performance analysis of semi-blind channel estimation, since the
channel coefficients are complex.
$E\left\{\delta\mathbf{d}_k\delta\mathbf{d}_k^T\right\}$ is also
necessary for the analysis in the next section. However, it is easy
to check that, for all $1\leq i,j\leq P^2$,
\begin{eqnarray}
\left(E\left\{\delta\mathbf{d}_k\delta\mathbf{d}^T_k\right\}\right)_{i,j}=\frac{1}{M}\left(\Sigma^k_{\delta\mathbf{d}}\right)_{i,\left(\begin{small}\mbox{mod}\end{small}(j,P)-1\right)P+\left\lceil\frac{j}{P}\right\rceil}\nonumber.
\end{eqnarray}
Then, it is further easy to check that
$\left(E\left\{\delta\mathbf{d}_k\delta\mathbf{d}^T_k\right\}\right)_{i,i}=0$,
$i=1,...,P^2$, except for
$\left\lceil\frac{i}{P}\right\rceil=\mbox{mod}(i,P)$ which means
that $\left(\mathbf{d}_k\right)_i$ are the diagonal elements of
$\mathbf{g}\mathbf{g}^H$, and thus are real. Therefore, we can
conclude that the real and imaginary parts of the SOS estimation
error are uncorrelated and have identical variances.

\itwsection{Performance of Semi-blind Channel Estimation} In this
section, we consider two approaches to semi-blind channel
estimation, namely moment-matching-based and subspace-based
algorithms. In semi-blind channel estimation, $M_t$ training symbols
are available (thus, the number of information symbols is $M-M_t$)
and the corresponding received training signal, which is represented
by a $NM_t$-vector, is given by
\begin{eqnarray}\label{Sig_with_train}
\mathbf{r}_t=\sum_{k=1}^KC_t(k)\mathbf{g}_k+\mathbf{n}_t,
\end{eqnarray}
where, similarly to (\ref{SignalModel}), $C_t(k)$ is the Sylvester
matrix constructed by
$$\tilde{\mathbf{s}}_k=\left(\left(\mathbf{s}_k^{(1)}x_k(1)\right)^H,...,\left(\mathbf{s}_k^{(M_t)}x_k(M_t)\right)^H\right)^H.$$

By multiplying $\frac{1}{M_t}C_t^T(k)$ on both sides of
(\ref{Sig_with_train}), the MAI for user $k$ can be completely
cancelled, as $M_t\rightarrow\infty$ since
$\frac{1}{M_t}C_t^T(k)C_t(k)\rightarrow\mathbf{I}$. Therefore, a
sufficient statistic for $\mathbf{g}_k$ is given by
$\bar{\mathbf{g}}_k\triangleq \frac{1}{M_t}C_t^T(k)\mathbf{r}_t$.
Thus, the equivalent training signal for user $k$ is given by
\begin{eqnarray}\label{train_noise}
\bar{\mathbf{g}}_k=\mathbf{g}_k+\tilde{\mathbf{n}}_k,
\end{eqnarray}
where $\tilde{\mathbf{n}}_k=\frac{1}{M_t}C_t^T(k)\mathbf{n}_t$ is
additive white complex Gaussian noise with variance
$\frac{\sigma_n^2}{M_t}$. As $M_t\rightarrow\infty$,
$\left(\tilde{\mathbf{n}}_k\right)_l$ become mutually independent
with respect to $k$ and $l$.

Since both $\left\{\bar{\mathbf{g}}_k\right\}$ and
$\left\{{\mathbf{d}}_k\right\}$ across different users are weakly
correlated when $M$ is sufficiently large , we can carry out the
channel estimation for each user separately, thereby reducing
considerably the computational complexity at cost of marginal
performance loss. For notational simplicity, we drop the subscripts
of user index throughout this section. The elements in $\mathbf{g}$
and $\mathbf{d}$ are denoted by $g_p$, $p=1,...,P$, and $d_p$,
$p=1,...,P^2$, respectively.

\subsection{Moment Matching}
\subsubsection{Moment-matching-based estimator} For applying the
asymptotically optimal estimator in ~\cite{Porat1993} and the
corresponding analysis, it is necessary to discuss the moment based
channel estimation in $\mathbb{R}$. For any complex vector
$\mathbf{v}$, we denote the corresponding real vector by
$\mathsf{v}\triangleq\left(\Re\left(\mathbf{v}\right)^T,\Im\left(\mathbf{v}\right)^T\right)^T$,
where $\Re\left(\mathbf{v}\right)$ and $\Im\left(\mathbf{v}\right)$
represent the real and imaginary parts of $\mathbf{v}$.

It should be noted that there are only $\frac{P(P+1)}{2}$ free
variables in $\hat{\mathbf{d}}$ and $P^2$ free variables in
$\hat{\mathsf{d}}$ (note that there are $P$ real elements in
$\hat{\mathbf{d}}$, corresponding to the diagonal elements in
$\mathbf{g}\mathbf{g}^T$) since $\hat{\mathbf{d}}$ is the
vectorization of a Hermite matrix. We denote the real vector of
these free variables by $\hat{\mathsf{d}}_f$, where the subscript
\textit{f} means \textit{free}.

On defining the observation vector
$\hat{\mathsf{z}}\triangleq\left(\bar{\mathsf{g}}^T,\hat{\mathsf{d}}_f^T\right)^T$,
$\mathsf{z}\triangleq\left(\mathsf{g}^T,\mathsf{d}_f^T\right)^T$ and
$\delta\mathsf{z}\triangleq\hat{\mathsf{z}}-\mathsf{z}$, the
asymptotically optimal estimator using estimates of the first and
second moments \cite{Porat1993} is obtained by minimizing a cost
function, which is given by
\begin{eqnarray}\label{OptEstimator}
J_{{opt}}\left(\hat{\mathbf{g}}\right)=\left(
\tilde{\mathsf{z}}\left(\hat{\mathbf{g}}\right)-\hat{\mathsf{z}}\right)^T\left(\Sigma_{\delta{\mathsf{z}}}\right)^{-1}\left(\tilde{\mathsf{z}}\left(\hat{\mathbf{g}}\right)-\hat{\mathsf{z}}\right),
\end{eqnarray}
where $\Sigma_{\delta{\mathsf{z}}}\triangleq
E\left\{\delta{\mathsf{z}}\delta{\mathsf{z}}^T\right\}$ and the
function $\tilde{\mathsf{z}}$ maps $\mathbf{g}$ to $\mathsf{z}$.

It is easy to check that
$$
\Sigma_{\delta \mathsf{z}}=\left(
\begin{array}{ll}
\frac{\sigma_n^2}{2M_t}\mathbf{I}_{2P\times 2P} & 0\\
0 & \frac{1}{M-M_t}\Sigma_{\delta \mathsf{d}_f}
\end{array}
 \right),
$$
where the covariance matrix $\Sigma_{\delta \mathsf{d}_f}$ can be
obtained from Prop. \ref{PropSOSMSE}.

However, due to the conclusion of Prop. \ref{PropSOSMSE},
$\Sigma_{\delta \mathsf{d}_f}$ is dependent on the realization of
$\mathbf{g}$, thus being unknown to the estimator. Therefore, the
optimal estimator is infeasible for practical applications. Since
the cross correlation of different elements in $\mathbf{d}_f$ is
small for moderate or large $\beta$, we can assume that
$\Sigma_{\delta \mathsf{d}_f}=\sigma_d^2\mathbf{I}$, thus resulting
in the cost function given by
\begin{eqnarray}\label{CostFunMomentMatch}
J\left(\hat{\mathbf{g}}\right)=w\left\|\mbox{vec}\left(\hat{\mathbf{g}}\hat{\mathbf{g}}^H\right)-\hat{\mathbf{d}}\right\|^2+(1-w)\left\|\hat{\mathbf{g}}-\bar{\mathbf{g}}\right\|^2,
\end{eqnarray}
where the weighting factor
$w=\frac{(1-\alpha)\sigma_n^2}{(1-\alpha)\sigma_n^2+\alpha\sigma_d^2}$.
Since minimizing (\ref{CostFunMomentMatch}) is equivalent to
obtaining $\hat{\mathbf{g}}$, which optimally matches the first and
second moment estimates in $\hat{\mathsf{z}}$, we call it
\textit{moment matching} based channel estimation.

\subsubsection{Performance analysis} Practically, the cost function
(\ref{CostFunMomentMatch}) can be minimized with iterative
optimization methods. For theoretical analysis, the optimal
$\mathbf{g}$ can also be obtained by taking derivatives of the cost
function (\ref{CostFunMomentMatch}) with respect to
$\hat{\mathsf{g}}$, resulting in $2P$ equations denoted by $F_i=0$,
$i=1,...,2P$. These equations determine a mapping ${\bf{\Psi}}$ from
the observation $\hat{\mathsf{z}}$ to $\hat{\mathsf{g}}$ in a
neighborhood of $\mathsf{z}$. Thus, by applying the implicit
function theorem \cite{Rudin1976}, we can obtain the Jacobian matrix
of ${\bf{\Psi}}$ at $\mathsf{z}$, which is given by
\begin{eqnarray}
\frac{\partial {\bf{\Psi}}}{\partial
\hat{\mathsf{z}}}\bigg|_{\hat{\mathsf{z}}=\mathsf{z}}=-\left(\frac{\partial
\mathbf{F}}{\partial\hat{\mathsf{g}}}\right)^{-1}\bigg|_{\hat{\mathsf{g}}=\mathsf{g}}\frac{\partial
\mathbf{F}}{\partial
\hat{\mathsf{z}}}\bigg|_{\hat{\mathsf{z}}=\mathsf{z}},
\end{eqnarray}
where $\mathbf{F}\triangleq \left(F_1,...,F_{2P}\right)^T$, provided
that $\frac{\partial \mathbf{F}}{\partial\hat{\mathsf{g}}}$ is
non-singular.

Therefore, for the channel estimation error $\delta
\mathsf{g}=\hat{\mathsf{g}}-\mathsf{g}$, we have the following
proposition, whose proof is essentially the same as that of Theorem
3.16 in ~\cite{Porat1993} and is omitted in this paper.
\begin{prop}\label{PropIV1}
In moment-matching-based semi-blind channel estimation, if
$\sqrt{M}\delta \mathsf{z}$ converges weakly to a random vector with
zero mean and a covariance matrix $\Sigma_{\delta \mathsf{z}}$ as
$M\rightarrow\infty$, then $M\delta \mathsf{g}$ converges weakly to
a random vector with zero mean and covariance matrix given by
\begin{eqnarray}
\Sigma_{\delta \mathsf{g}}=\left(\frac{\partial
{\bf{\Psi}}}{\partial \hat{\mathsf{z}}}\Sigma_{\delta
\mathsf{z}}\left(\frac{\partial {\bf{\Psi}}}{\partial
\hat{\mathsf{z}}}\right)^T\right)\bigg|_{\hat{\mathsf{z}}=\mathsf{z}},
\end{eqnarray}
provided that the Jacobian matrix $\frac{\partial
{\bf{\Psi}}}{\partial
\hat{\mathsf{z}}}\big|_{\hat{\mathsf{z}}=\mathsf{z}}$ is
nonsingular.
\end{prop}

\subsubsection{Performance bound} Using Lemma 3.1 in
\cite{Porat1993}, the asymptotic performance of SOS based blind
channel estimation can be lower bounded by that of the
asymptotically optimal estimator, which is given by
\begin{eqnarray}\label{lowerbound}
\Sigma_{\delta \mathsf{g}}\geq \left(\left(\frac{\partial
\hat{\mathsf{z}}}{\partial
\hat{\mathsf{g}}}\right)^{T}\Sigma_{\delta
\mathsf{z}}^{-1}\frac{\partial \hat{\mathsf{z}}}{\partial
\hat{\mathsf{g}}}\right)^{-1}\bigg|_{\hat{\mathsf{g}}=\mathsf{g}},
\end{eqnarray}
where
\begin{eqnarray}
\frac{\partial \hat{\mathsf{z}}}{\partial
\hat{\mathsf{g}}}\bigg|_{\hat{\mathsf{g}}=\mathsf{g}}=\left(\begin{array}{ll}
\mathbf{I}_{2P\times 2P}\\
\frac{\partial \hat{\mathsf{d}}_f}{\partial
\hat{\mathsf{g}}}\bigg|_{\hat{\mathsf{g}}=\mathsf{g}}
\end{array}\right),
\end{eqnarray}

\subsection{Subspace-based Approach}
\subsubsection{Subspace-based estimator} Another methodology for
semi-blind channel estimation is to make use of the subspace
estimated from the SOS estimates. In this section, we adopt a simple
subspace algorithm, which can be carried out in the following three
steps:
\begin{enumerate}

\item The one-dimensional subspace spanned by $\mathbf{g}$ is
obtained from the unit eigenvector $\mathbf{u}$ of matrix
$\mbox{vec}^{-1}\left(\hat{\mathbf{d}}\right)$
\footnote{$\mbox{vec}^{-1}$ is the inverse operation of
$\mbox{vec}$, thus $\mbox{vec}^{-1}\left(\hat{\mathbf{d}}\right)$ is
an estimation of matrix $\mathbf{g}\mathbf{g}^H.$} corresponding to
its largest eigenvalue.

\item The channel estimate from the training symbols, namely
$\bar{\mathbf{g}}$, is projected onto the subspace, thus decreasing
the noise power and obtaining a tentative channel estimate, which is
given by
\begin{eqnarray}\label{25}
\tilde{\mathbf{g}}={\mathbf{u}}^H\bar{\mathbf{g}}{\mathbf{u}}.
\end{eqnarray}

\item The final channel estimate is obtained by combining
$\bar{\mathbf{g}}$ and $\tilde{\mathbf{g}}$ linearly:
\begin{eqnarray}
\hat{\mathbf{g}}=\omega\tilde{\mathbf{g}}+(1-\omega)\bar{\mathbf{g}},\nonumber
\end{eqnarray}
where $\omega$ is a weighting factor, which will be optimized later.
\end{enumerate}

This algorithm can also be regarded as tackling the phase ambiguity
incurred by SOS based blind channel estimation by making use of the
training symbols.

\subsubsection{Perturbation on signal subspace} On denoting the
acute angle between the subspace spanned by $\mathbf{g}$ and the
unit eigenvector ${\mathbf{u}}$ by
$\theta=\arccos\left(\frac{\left|\mathbf{g}^H{\mathbf{u}}\right|}{\left\|\mathbf{g}\right\|}\right)$,
we can obtain an explicit asymptotic expression for $E\left\{\sin
^2\theta\right\}$, which is given in the following Proposition.
\begin{prop}~\label{PropIV3}
As $K,N,M\rightarrow\infty$,
\begin{scriptsize}
\begin{eqnarray}\label{27}
\sigma_\theta^2&\triangleq & ME\left\{\sin
^2\theta\right\}\nonumber\\ &\rightarrow&
\frac{(P-1)\left(\left(2\beta+\|\mathbf{g}\|^2\right)\sigma_n^2+\left(\sigma_n^2\right)^2+\beta^2+\frac{2\beta}{P}\right)}{\|\mathbf{g}\|^4}.
\end{eqnarray}
\end{scriptsize}
\end{prop}

\subsection{Performance of channel estimation} On defining
$\delta{\mathbf{u}}\triangleq
\frac{-\mathbf{g}+{\mathbf{u}}^H\mathbf{g}{\mathbf{u}}}{\|\mathbf{g}\|}$,
(\ref{25}) can be rewritten as
\begin{eqnarray}
\tilde{\mathbf{g}}=\frac{\|\mathbf{g}\|^2}{\left|\mathbf{u}^H\mathbf{g}\right|^2}\left(\frac{\mathbf{g}}{\|\mathbf{g}\|}+\delta{\mathbf{u}}\right)^H\left(\mathbf{g}+\tilde{\mathbf{n}}\right)\left(\frac{\mathbf{g}}{\|\mathbf{g}\|}+\delta{\mathbf{u}}\right),\nonumber
\end{eqnarray}
where $\tilde{\mathbf{n}}$ is defined in (\ref{train_noise}). It is
easy to check that $\delta {\mathbf{u}}^H\delta
{\mathbf{u}}\approx\sin^2\theta$, which becomes tight as
$\theta\rightarrow 0$. We assume that the different elements in
$\delta {\mathbf{u}}$ are mutually independent, and the real and
imaginary parts of each element are also mutually independent with
identical variances. These assumptions can be validated by
simulation.

By ignoring the higher order terms and applying
$\left|\mathbf{u}^H\mathbf{g}\right|^2\rightarrow\|\mathbf{g}\|^2$,
we can obtain
\begin{scriptsize}
\begin{eqnarray}\label{sigma_g_subspace}
\sigma_g^2&\triangleq &
\frac{ME\left\{\delta\mathbf{g}^H\delta\mathbf{g}\right\}}{P}\nonumber\\
&=&\frac{\omega^2\|\mathbf{g}\|^2\left(1+\frac{1}{P}\right)\sigma_\theta^2}{(1-\alpha)P}+\frac{(1-\omega)^2(P-1)\sigma_n^2}{P\alpha}+\frac{\sigma_n^2}{P\alpha},
\end{eqnarray}
\end{scriptsize}
where the optimal $\omega$ is given by
\begin{eqnarray}
\omega_{opt}=\frac{\frac{(P-1)\sigma_n^2}{\alpha}}{\frac{\|\mathbf{g}\|^2\left(1+\frac{1}{P}\right)\sigma_\theta^2}{1-\alpha}+\frac{(P-1)\sigma_n^2}{\alpha}}.
\end{eqnarray}

When $\omega=0$, $\sigma_g^2=\frac{\sigma_n^2}{\alpha}$, which
equals the performance of training-symbol-based estimation. Thus,
the subspace-based channel estimation with optimal $\omega$ surely
attains better performance than training symbol based estimation.
When the subspace estimation is perfect and $\omega=1$,
$\sigma_g^2=\frac{\sigma_n^2}{\alpha P}$, which means that noise
level is lowered by a factor of $\frac{1}{P}$. Therefore, the last
term in (\ref{sigma_g_subspace}) is the performance with perfect SOS
estimation and the first term in (\ref{sigma_g_subspace}) represent
the penalty incurred by imperfect SOS estimation.

\itwsection{Simulation results}
\subsection{Moment-matching-based
Channel Estimation}
We define the \textit{blind estimation
efficiency} as
\begin{eqnarray}
\eta\triangleq
\frac{\frac{\sigma_n^2}{\sigma_g^2}-\alpha}{1-\alpha}.
\end{eqnarray}
Intuitively, $\eta$ quantifies how many training symbols each
information symbol is equivalent to for the purpose of semi-blind
channel estimation. Figure 1 shows the efficiency $\eta$ obtained
from Prop. \ref{PropIV1} versus $\beta$ and $\sigma_n^2$ when $P=3$
and $\alpha=0.2$. We observe that $\eta$ decreases monotonically
with $\beta$ since larger $\beta$ implies more MAI for SOS
estimation. An interesting observation is that the efficiency is
small for both high and low noise levels and achieves a maximum
(around 0.3) for moderate noise power (around $\sigma_n^2=1$). An
intuitive explanation is that, when $\sigma_n^2$ is large, the SOS
estimation performance is poor since the SOS error variance contains
a term proportional to $\left(\sigma_n^2\right)^2$; when
$\sigma_n^2$ is small, the SOS estimation error is dominated by the
MAI, thus achieving much lower efficiency than training-symbol-based
estimation.

Figure 2 shows the efficiency versus $P$ and $\alpha$ when
$\sigma_n^2=0.5$ and $\beta=0.5$. An important observation from this
figure is that, for large $\alpha$ and $P$, the efficiency $\eta<0$,
which means that it is harmful to incorporate the information
symbols into the channel estimation. This arises from the
suboptimality of the channel estimator in (\ref{CostFunMomentMatch})
since the exact covariance matrix of the SOS estimation error is
unknown and the estimated weighting factor $w$ is also suboptimal.
Thus, the unreliable SOS estimation may have negative impact on the
reliable estimation from the training symbols, when $\alpha$ is
large. When $\alpha$ is small, the efficiency achieves its maximum
for small $P$; however, the efficiency does not change much with
respect to $P$.

From the simulation results of Fig. 1 and Fig. 2, we can draw a
conclusion that moment-matching-based channel estimation is suitable
for systems with small system load, small numbers of training
symbols, small channel order and moderate noise power.

\subsection{Subspace-based Channel Estimation}
The blind estimation efficiency is shown in Fig. 3 (versus $\beta$
and $\sigma_n^2$ when $\alpha=0.1$ and $P=3$) and in Fig. 4 (versus
$\alpha$ and $P$ when $\beta=0.5$ and $\sigma_n^2=0.5$). Similar to
that of moment-matching-based channel estimation, the optimal
efficiency is attained at moderate noise levels (around
$\sigma_n^2=1$), and the efficiency decreases with respect to
$\beta$. However, unlike moment-matching-based channel estimation,
the efficiency of subspace-based channel estimation increases with
$\alpha$ and $P$. Thus, subspace-based channel estimation is
suitable for systems with large channel order and large numbers of
training symbols, which are opposite to the desirable conditions for
moment-matching-based channel estimation. Also, we see that the
blind estimation efficiency of subspace-based channel estimation is
always positive, indicating that this technique is more robust than
moment-matching-based channel estimation.

\itwsection{Conclusions} In this paper, we have analyzed the
performance of SOS based semi-blind channel estimation in long-code
CDMA systems. The main results include the following:
\begin{itemize}
\item An explicit expression for the covariance matrix of the SOS
estimation error has been obtained in the large system limit within
some assumptions.

\item Expressions for the performance of two types of semi-blind
channel estimation have been obtained. Particularly, we have
obtained an asymptotic expression characterizing the perturbation of
the eigenvectors of covariance matrices, which can be applied to
other problems as well.

\item The blind estimation efficiency has been obtained from
simulation results, which show that the SOS based semi-blind channel
estimation attains high efficiency for systems operating in the
moderate noise region. Moment-matching-based estimation is suitable
for systems with small channel order and small numbers of training
symbols, while subspace-based estimation achieves good performance
in systems with large channel order and large numbers of training
symbols.
\end{itemize}

\begin{figure}
  \centering
  \includegraphics[scale=0.4]{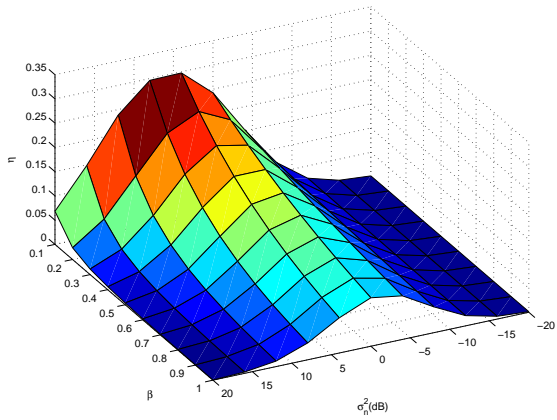}
  \caption{Blind channel estimation efficiency versus noise power and system load in moment-matching-based channel estimation}\label{}
\end{figure}

\begin{figure}
  \centering
  \includegraphics[scale=0.4]{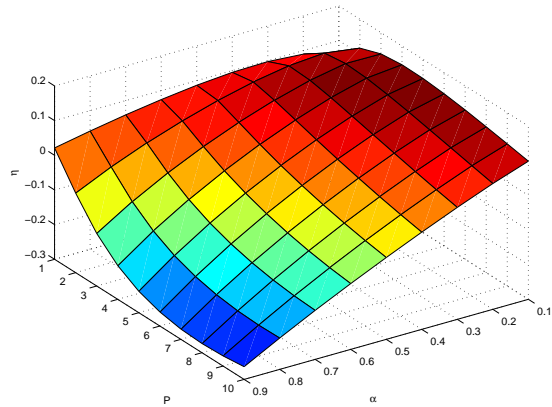}
  \caption{Blind channel estimation efficiency versus channel order and training symbol proportion in moment-matching-based channel estimation}\label{}
\end{figure}

\begin{figure}
  \centering
  \includegraphics[scale=0.4]{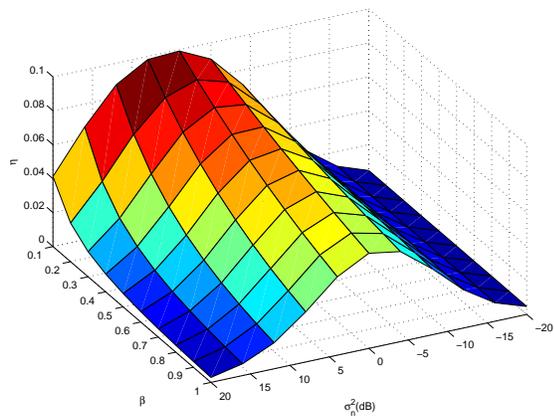}
  \caption{Blind channel estimation efficiency versus noise power and system load in subspace-based channel estimation}\label{}
\end{figure}

\begin{figure}
  \centering
  \includegraphics[scale=0.4]{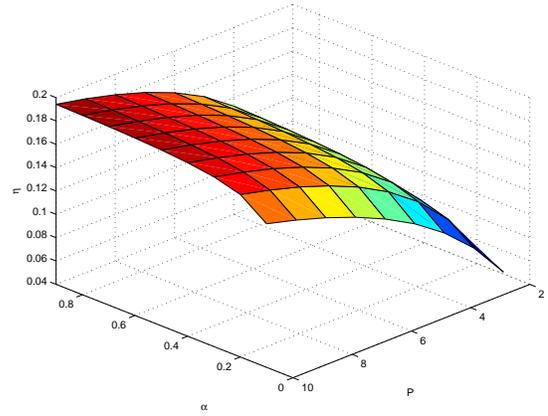}
  \caption{Blind channel estimation efficiency versus channel order and training symbol proportion in subspace-based channel estimation}\label{}
\end{figure}

\end{itwpaper}

\begin{itwreferences}
\bibitem{Bensley1996}
S. E. Bensley and B. Aazhang, ``Subspace-based channel estimation
for code division multiple acess communication systems,'' {\em IEEE
Trans. Commun.}, Vol.~44,
  pp.~1009-1020, Aug. 1996.

\bibitem{Buchoux2000}
V. Buchoux, O. Cappe, E. Moulines and A. Gorokhov, ``On the
performance of semi-blind subspace-based channel estimation,'' {\em
IEEE Trans. Signal Processing}, Vol.~48,
  pp.~1750-1759, June 2000.

\bibitem{Buzzi2001}
S. Buzzi and H. V. Poor, ``Channel estimation and multiuser
detection in long-code DS/CDMA systems,'' {\em IEEE J. Select. Areas
Commun.}, Vol.~19, pp.~1476-1487, Aug. 2001.

\bibitem{Hassibi2003}
B. Hassibi and B. M. Hochwald, ``How much training is needed in
multiple-antenna wireless links?,'' {\em IEEE Trans. Inform.
Theory}, Vol.~49,
  pp.~951-963, Apr. 2003.

\bibitem{LiHu2004}
H. Li and H. V. Poor, ``Impact of channel estimation error on
multiuser detection via the replica method ,'' {\em Proc. 2004 IEEE
Global Telecomm. Conf.}, Dallas, TX, Nov. 30 - Dec. 2, 2004.

\bibitem{LiHuPreprint}
H. Li and H. V. Poor, ``Performance analysis of semi-blind channel
estimation in fading DS-CDMA channels,'' submitted to {\em IEEE
Trans. Signal Process.}.

\bibitem{Liu1996}
H. Liu, G. Xu, L. Tong and T. Kailath, ``Recent developments in
blind channel equalization: From cyclostationarity to subspaces,''
{\em Signal Processing}, Vol.~50, pp.~83-99, Jan. 1996.

\bibitem{Porat1993}
B. Porat, {\em Digital Processing of Random Signals}.
\newblock Prentice Hall, Upper Saddle River, NJ, 1994.

\bibitem{Rudin1976}
W. Rudin, {\em Principles of Mathematical Analysis}.
\newblock McGraw-Hill Inc., New York, NY, 1976.

\bibitem{Sun2001}
J. Sun, {\em Analysis of Matrix Perturbation}.
\newblock Science Press, Beijing, China, 2001.

\bibitem{ZhengyuanXu2000}
Z. Xu and M. K. Tsatsanis, ``Blind channel estimation for long code
multiuser CDMA systems,'' {\em IEEE Trans. Signal Processing},
Vol.~48,
  pp.~988-1001, Apr. 2000.

\bibitem{ZhengyuanXu2004}
Z. Xu, ``Effects of imperfect blind channel estimation on
performance of linear CDMA receivers,'' {\em IEEE Trans. Signal
Processing}, Vol.~52,
  pp.~2873-2884, Oct. 2004.

\bibitem{Zeng1997}
H.~H. Zeng and L.~Tong, ``Blind channel estimation using the second
order statistics: Algorithms,'' {\em IEEE Trans. Signal Processing},
Vol.~45,
  pp.~1919-1930, Aug. 1997.

\end{itwreferences}
\end{document}